Scripps Research Translational Institute

# Exploring the Synchrony Between Body Temperature and HR, RR, and Aortic Blood Pressure in Viral/Bacterial Disease Onsets with Signal Dynamics

*Correlational Study*

Camille Dunning

*Scripps Research Translational Institute, La Jolla, CA*
Submitted 24 September 2020

## Abstract

Signal-based early detection of illnesses has been a key topic in research and hospital settings; it reduces technological costs and paves the way for quick and effective patient-care operations. Elementary machine learning and signal processing algorithms have proven to be sufficient in classifying the onset of viral and bacterial conditions before clinical symptoms are shown. Inspired by these recent developments, this project employs signal dynamics analysis to infer changes in vital signs (temperature, respiration, and heart rate). The results demonstrate that the trends of one vital function can be predicted from that of another. In particular, it is shown that heart rate and respiration typically change shortly after body temperature, and aortic blood pressure follows. This is not an etiologically specific approach, but if advanced further, it can enable patients and wearable system users to tame these changes and prevent immediate symptoms.

*Keywords:* Exploratory data analysis, cardiology, body temperature, heart rate

## Introduction

Recent medical research has witnessed a growing demand for technology to detect pathogen onsets earlier than traditional biomolecule based methods. This paper is inspired by new machine learning based procedures that classify fever onsets by catching quantifiable anomalies in high-resolution physiological time series data. The focus of these processes was to detect diseases before symptoms become overt using binary classifiers over a threshold. This project seeks to employ similar electrocardiographic tracings, but rather than zero in on specific pathogens, explore mathematical techniques to infer changes in the vital signs, heart rate/RR wavelengths and aortic blood pressure (AOPA). This paper first provides an overview of the preprocessing techniques and missing data handling using mean imputation, and then the calculation of time-lagged cross-correlation (TLCC) of three tracings, HR, RR, and AOPA, against body temperature only after viral or bacterial exposure. The dataset description is in the next section of this paper. Based on the acquired cross-correlation signal, an ideal "offset" is calculated to show how many frames ($fps = 48$), one of the three signals, $y_A$, must be shifted such that it correlates highest with Temperature $y_B$. A negative offset of $y_B$ from $y_A$ indicates that it lags behind $y_A$, and a positive offset means it precedes $y_A$.

This study finds that synchrony-based analysis of vital sign signals, provided that these signals are high enough resolution, can uncover sequential patterns and lead to algorithms for wearable sensor systems to predict changes in vital signs.

## Methods

*Data from Animal Studies*

Data were acquired from a study (Milechin et. al., 2017) that designed a classification method to recognize pathogen-induced illnesses in a non-human primate subject based on quantitative abnormalities in its corresponding physiological waveforms. Groups of non-juvenile non-human primates—cynomolgus macaques, African green monkeys, and rhesus macaques—were transferred to containment (Janosko et. al., 2016) 5 to 7 days before exposure, and pre-exposure data were collected 4 to 6 before. While under sedation, the NHPs were exposed to one of a set of viral diseases (Ebola, Marburg, Lassa, and Nipah viruses) or the bacterial disease, *Y. pestis*, via one of the following: aerosol, intramuscular injection, or intratracheal instillation. The viral isolate stocks were provided by the United States Army Medical Research Institute of Infectious Diseases (USAMRIID) from various outbreaks.

*Data Preprocessing and Transformation*

The physiological waveform data come in the form of sequential time-series, and were conditioned such that instrumental noise and artifacts are eliminated. The statistics displayed in the datasets were computed in 30-minute intervals, to offset computational deprivation and analytical performance, and standardized to remove short-term variations from diurnal rhythms. Standardization was accomplished with the following: $(x_i - \mu_i)/\sigma_i$, for the $i^{th}$ interval, where $x_i$ is the corresponding data point, and μ and σ are the respective interval mean and standard deviation.

Loss of statistical power and precision in cohort studies comes at the heels of missing data, which, in the Milechin et. al. study, came mostly from differences in animal and sedation instrument handling. The variation among the six exposure studies in missing data ratios is shown in Figure 1, with the subject's study exposure cohort shown as a dashed blue line.

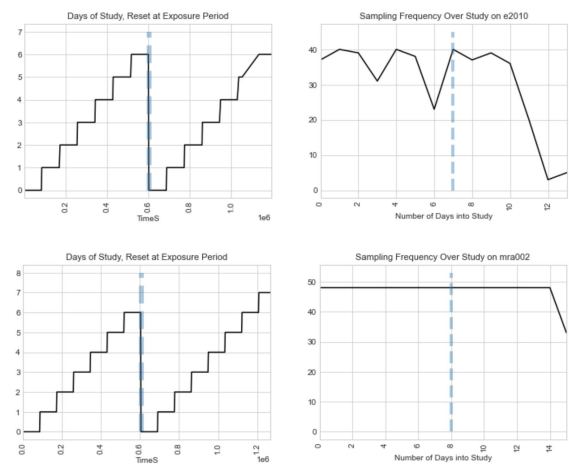

**Figure 1:** The physiological data from subject e2010 in the Ebola study on cynomolgus macaques exposed via aerosol had several gaps of missingness, and 48 aggregated (mean) samples are never collected in a day. The data from subject mra002 in the Marburg study on rhesus macaques exposed via aerosol shows consistent sampling with 48 aggregates per day except at the end of the subject's study. 48 data points should be collected to fill a 24h time period with 0.5h intervals.

Mean imputation (Zhang, 2016), or mean value substitution mitigates incompleteness in the independent datasets. Mean imputation maintains proportionality in sample sizes but runs the risk of lower variability, and less lossy interpolation methods for high-resolution time series may be considered in future proceedings. Figure 2 displays the results of mean imputation on subjects e1001 and e1009, wherein the red gaps on the left were filled with the daily mean.

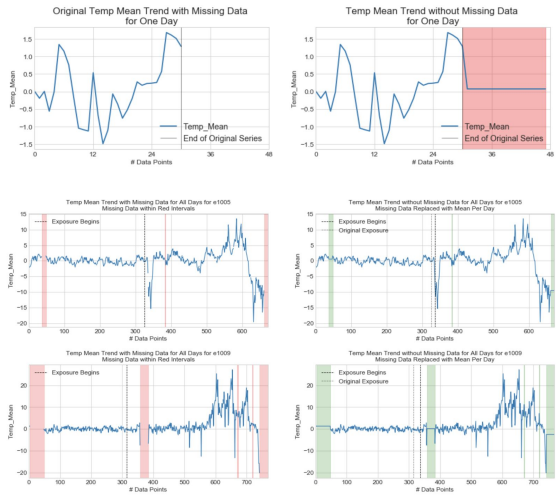

**Figure 2:** The aggregated ECG temperature waveform's gaps were detected and connected by the mean for the range's corresponding day, thereby shifting the exposure cohort as well as increasing the subject's sample size. An example for a single day is shown at the top of the figure. Finally, the structure of each study was transformed to adhere to a hierarchy, wherein time-series were separated individually while maintaining relationships and trends across one study. 1d/48-sample moving averages and standard deviations were generated for the Temperature, Respiration Rate, Aortic Blood Pressure (mmHg), Heart Rate, and RR Interval Length. These features were represented as a DataFrame Index per subject, wrapped by a subject DataFrame Index. This was accomplished with a MultiIndex, the hierarchical analogue of the Standard Index, both of which are available in the pandas package for Python. With multiple time series in one dataset, a hierarchical approach is preferred to streamline iterative analysis. These transformed hierarchical studies and their contained rolling average and standard deviation trends were used for the rest of this study.

*Cross-Correlation*

For each study, pre-exposure and post-exposure periods for each subject were separated and cross-correlation was performed in parallel. Cross-correlation seeks to quantify the synchrony of two series as a function of the time lag between one another, and has the potential to reveal hidden sequences. This project utilized NumPy's cross-correlation capabilities; SciPy provides a similar function in its signal processing package. In the cross-correlation procedure, two signals, $f$ and $g$, are objectively compared by shifting $g$ only, and the lag between $f$ and $g$ at which the correlation coefficient is the highest is returned. Methodically, we seek to estimate a delay $D$ with $\tau = \widehat{D}$ for which $\varphi(\tau)$ is maximized:

$$\varphi(\widehat{D}) = max(\varphi(\tau)) = max \int_{-T/2}^{T/2} f(t)g(t+\tau)dt, \quad \text{provided}$$

that $f$ and $g$ are periodic functions of period $T$.

NumPy's correlation function returns a subset of the discrete linear cross-correlation of two input signals, in this case, the Temperature, and any of the Respiration Rate, Aortic Blood Pressure, HR, and RR Length. The mode is set to return a centered CC signal of length max(signal A, signal B). This resulting third signal was divided by the length of signal A for convenient scaling, and the optimal offset is calculated, mainly to account for centering, by the following: $argmax(CC\ signal) - ceil(|A|/2)$. Note that since the RR interval lengths are inversely correlated with the Heart Rate, their curve had to be flipped vertically when cross-correlating with Body Temperature.

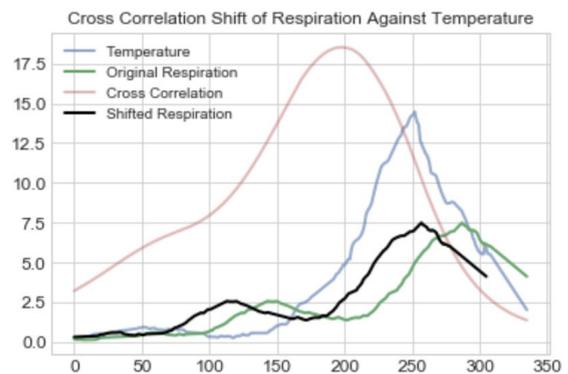

**Figure 3:** The ideal offset occurs when a signal, respiration-vs.-time, for example, is shifted from its original position such that the correlation between another stationary signal, temperature-vs.-time, is maximized. The signal is shifted by a number of frames that is calculated by the above formula. This results in the shifted signal being centered; note that this did not affect the position of the CC curve in red.

## Results

The majority of the signals from the post-exposure physiological data lag behind the Temperature signal for their respective disease. Although, there are instances, particularly found in Nipah-infected primates, in which these trends precede the Temperature. Note that there are less data points available for the mean respiration: respiration waveforms were not collected from the Nipah cases.

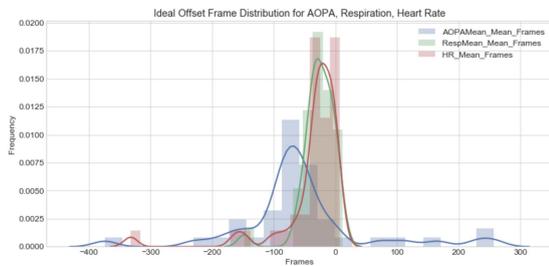

**Figure 4:** Aortic blood pressure changes last out of all physiological waveforms, with there being the numbers for cross correlation shifts being predominantly negative. The lagging trend also holds true for the Respiration and Heart Rate signals, and their frame-offset distributions are similar in that they rarely foreshadow fluctuations in body temperature (i.e., there are few recorded positive offsets). There may be factors for the resemblance between the HR and Respiration frame distributions, but examining such was beyond the scope of this project, and the sample sizes were different. There is more variability in frame-offset distribution for AOPA. The lead-lag measurements against Temperature for the three signals, plus mean RR interval lengths, are documented in Table 1.

| Ideal Frames | AOPA | Resp. | HR | RR |
|---|---|---|---|---|
| Mean | -56.67 | -27.28 | -32.28 | -38.07 |
| Standard Dev. | 107.01 | 27.8 | 68.38 | 58.34 |
| Min | -374 | -148 | -332 | -328 |
| Max | 257 | 11 | 7 | 191 |

**Table 1:** While AOPA's lag is the lowest on average, it also is the most variable. It is preceded by RR and HR, the two of whose lag differences are simply explained by the independence of the two signals when measuring cross correlation, even though they are dependent on their own (recall their inverse correlation: lower HR corresponds to bigger RR intervals). The rise in Mean Respiration most immediately follows the rise in Temperature.

Certain diseases might yield special cases that shift the overall lag distribution for the type of signal.

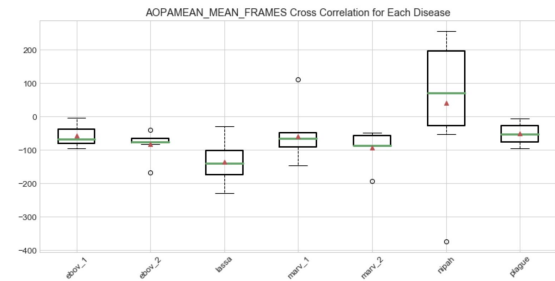

**Figure 5:** Nipah is the only disease in which aortic blood pressure precedes temperature. However, the AOPA signal for at least one of the Nipah subjects came behind temperature; the lag distribution is clearly the most variable out of those for all of the recorded diseases.

For all diseases, none of the average shifting factors exceeded zero. The method of gauging signal dynamics is intended to have low etiological specificity. It is informative of the onset of cardiological changes themselves rather than of particular pathogens.

## Discussion

The effects observed in this project help establish high body temperature as a symptom of high blood pressure. All NHPs were exposed to a fever as a result of the pathogens, and their body temperature and heart rates subsequently increased (Knibbs, 2019). The lower negative ideal offset for the heart rate compared to the aortic blood pressure also confirms this, as increased heart rates should heighten blood pressure.

Notwithstanding this sequential relationship of the vital signs, it is also understandable in some cases that blood pressure can precede heart rate. High blood pressure can be a precursor to heart rates passing safe levels as the arteries become increasingly resistant to blood flow and they have to toil to circulate blood ("Vital Signs"). The distribution sequence also verifies that the respiration rate increases with the onset of a fever.

Visualized properly, cross-correlation comparisons are a straightforward approach to evaluating lead-lag relationships in time series, and analyzing them is an important first step in advancing machine learning techniques and regression techniques on physiological for early pathogen detection.

## Conclusion

The sequence of increases in body temperature, HR and RR intervals, respiration, and aortic blood pressure after exposure to a viral or bacterial pathogen can be demonstrated by examining physiological data on its own. Elementary lag measurement methods can be sufficient to help predict when changes in one vital sign may come at the heels of the changes of another.

## Acknowledgements

I would like to thank my mentor, Professor Giorgio Quer, for his statistics, analytics, and visualization guidance. I would also like to thank Theresa Hill and Dr. Laura Nicholson of the Scripps Research Translational Institute for giving me the opportunity to participate in the Student Research Internship during which I undertook this project.